\documentclass[12pt,twoside]{article}

\usepackage{amsmath,amssymb}
\usepackage{epsfig}
\usepackage{slashed}
\usepackage{graphicx,color}
\usepackage[sort&compress]{natbib}

\bibpunct{[}{]}{,}{n}{}{,}

\newcommand{\bra}[1]{\ensuremath{\langle #1 |}}   
\newcommand{\ket}[1]{\ensuremath{| #1 \rangle}}   
\newcommand{\sprod}[2]{\ensuremath{\left\langle #1 |%
                     #2 \right\rangle}}  
\newcommand{\ev}[1]{\ensuremath{\left\langle #1 %
                     \right\rangle}} 

\renewcommand{\vec}[1]{{\mathbf{#1}}}
\newcommand{\MB}{M\"{o}ssbauer}

\renewcommand{\H}{{\rm H}}
\newcommand{\He}{{\rm He}}

\title{\MB\ neutrinos in quantum mechanics and quantum field theory}
\author{Joachim Kopp\footnote{Email: jkopp@mpi-hd.mpg.de} \\[0.2cm]
         {\it Max--Planck--Institut f\"{u}r Kernphysik}, \\
         {\it Postfach 10 39 80, 69029 Heidelberg, Germany}}
\date{29 April 2009}

\makeatletter
\def\@maketitle{%
     \renewcommand{\thefootnote}{\alph{footnote}}
     \newpage
     \vspace*{0.5em}
     \begin{center}%
     \let \footnote \thanks
       {\Large\bf \@title \par}%
       \vskip 1.0em%
       {\normalsize
         \lineskip .5em%
         \begin{tabular}[t]{c}%
           \@author
         \end{tabular}\par}%
       \vskip 0.7em%
       {\normalsize \@date}%
     \end{center}%
     \par
     \vskip 0.5em}
\makeatother

\makeatletter
\renewcommand\section{\@startsection {section}{1}{\z@}%
                                   {-3.5ex \@plus -1ex \@minus -.2ex}%
                                   {2.3ex \@plus.2ex}%
                                   {\normalfont\large\bfseries}}
\makeatother

\begin{document}

\maketitle

\begin{abstract}
  We demonstrate the correspondence between quantum mechanical and quantum
  field theoretical descriptions of \MB\ neutrino oscillations.  First, we
  compute the combined rate $\Gamma$ of \MB\ neutrino emission, propagation,
  and detection in quantum field theory, treating the neutrino as an internal
  line of a tree level Feynman diagram. We include explicitly the effect of
  homogeneous line broadening due to fluctuating electromagnetic fields in the
  source and detector crystals and show that the resulting formula for $\Gamma$
  is identical to the one obtained previously~\cite{Akhmedov:2008jn} for the
  case of inhomogeneous line broadening. We then proceed to a quantum
  mechanical treatment of \MB\ neutrinos and show that the oscillation,
  coherence, and resonance terms from the field theoretical result can be
  reproduced if the neutrino is described as a superposition of Lorentz-shaped
  wave packet with appropriately chosen energies and widths. On the other hand,
  the emission rate and the detection cross section, including localization and
  Lamb-\MB\ terms, cannot be predicted in quantum mechanics and have to be put
  in by hand.
\end{abstract}

\section{Introduction}

The possibility of exploiting the \MB\ effect in weak interactions to enhance
the small neutrino cross sections~\cite{Visscher:1959,Kells:1983,Kells:1984nm,
Raghavan:2005gn,Raghavan:2006xf} has recently received considerable interest,
both from the experimental side~\cite{Raghavan:2005gn,Raghavan:2006xf,
Potzel:2006ad,Potzel:2008xk} and from the theoretical side~\cite{Minakata:2006ne,
Bilenky:2006hk,Bilenky:2007vs,Akhmedov:2008jn,Bilenky:2008ez,Akhmedov:2008zz,
Bilenky:2008dk,Cohen:2008qb,Parke:2008cz}. In the proposed experiment,
neutrinos are emitted from $^3\H$ atoms embedded into a metal crystal and
absorbed by $^3\He$ atoms embedded into a similar crystal. With very optimistic
assumptions on the source activity (1~MCi), the fraction of recoilfree
emissions and absorptions (0.28 each), and the achievable spectral line width
($\Delta E/E \sim 10^{-11}\ \text{eV} / 18.6\ \text{keV} \sim 5 \cdot
10^{-16}$), it has been estimates that an event rate of $10^3$ per day could be
achieved for a detector containing 1~g of $^3\He$ and placed at a baseline $L =
10$~m~\cite{Raghavan:2006xf}.%
\footnote{It has been suggested recently that it might even be possible to
reach a line width of $\mathcal{O}(10^{-24}\ \text{eV})$, corresponding to the
natural line width of tritium
decay~\cite{Raghavan:2008cs,Raghavan:2008tb,Raghavan:2009hj}.  This would imply
an additional enhancement of the event rate by a factor of $10^{13}$, allowing
for smaller sources and detectors, or for longer baselines. However, the
arguments given in~\cite{Raghavan:2008cs,Raghavan:2008tb,Raghavan:2009hj} in
favor of this additional enhancement have been disproven in
ref.~\cite{Potzel:2009a}.}
These events could be counted by observing the subsequent decays of the
produced $^3\H$ in the detector, or by chemically extracting and counting the
number of produced $^3\H$ atoms.  However, it is far from clear whether the
above experimental performance can be achieved in practice, and, in fact, the
event rate may well be many orders of magnitude smaller~\cite{Potzel:2008xk},
so that the question of whether a \MB\ neutrino experiment can be realized in
practice is still open.

In spite of this, \MB\ neutrinos have already now proven to be an excellent
test case for studying the quantum mechanics and quantum field theory of
neutrino oscillations theoretically. In particular, their extremely small
energy spread of $\mathcal{O}(10^{-11}\ {\rm eV})$~\cite{Potzel:2006ad,
Coussement:1992a} has led to the question whether a coherent emission and
absorption of different neutrino mass eigenstates, which is a prerequisite for
oscillations, is possible~\cite{Bilenky:2007vs,Bilenky:2008dk}. Even though a
detailed quantum field theoretical treatment, requiring no a priori assumptions
on the neutrino wave function, shows that oscillations do occur in a \MB\
neutrino experiment~\cite{Akhmedov:2008jn,Akhmedov:2008zz}, such an approach
also reveals that \MB\ neutrinos are special because many of the assumptions
and approximations that are commonly made in the theoretical treatment of
conventional neutrino oscillation experiments are invalid for them.

In this paper, we will use the example of \MB\ neutrinos to discuss the
correspondence between quantum mechanical and quantum field theoretical
approaches to neutrino oscillations.  In sec.~\ref{sec:qft}, we will derive
the combined rate of \MB\ neutrino emission, propagation, and detection in
quantum field theory (QFT). We will for the first time explicitly include
homogeneous line broadening effects arising from fluctuating electromagnetic
fields in the solid state crystals forming the \MB\ source and detector, and we
will show that, as anticipated in ref.~\cite{Akhmedov:2008jn}, the result is
identical to the one obtained for inhomogeneous line broadening due to crystal
imperfections. We will then derive the same result from quantum mechanics (QM)
in sec.~\ref{sec:qm}, treating the neutrino as a wave packet.  A comparison of
the QFT and QM approaches will show that QM is inferior to QFT because more ad
hoc assumptions are required, e.g.\ on the shape and width of the neutrino wave
packets.  If, however, all parameters are chosen appropriately in the QM
formalism, the QFT result can be reproduced.  In sec.~\ref{sec:discussion},
we will discuss our results and conclude.

\section{\MB\ neutrinos in quantum field theory and homogeneous line broadening}
\label{sec:qft}

\begin{figure}
  \begin{center}
    \includegraphics{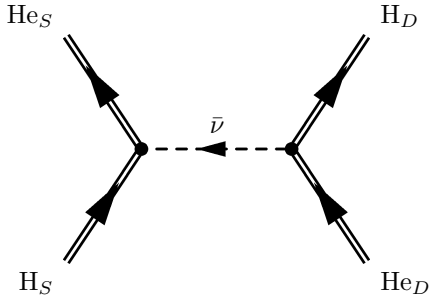}
  \end{center}
  \caption{Feynman diagram for neutrino emission and absorption
           in the $^3$H\,--\,$^3$He system.}
  \label{fig:feyn}
\end{figure}

To compute the amplitude for \MB\ neutrino production, propagation, and
detection in QFT, we follow the formalism developed in~\cite{Akhmedov:2008jn}
and consider the Feynman diagram shown in fig.~\ref{fig:feyn}. Here, the
external lines correspond to the $^3\H$ and $^3\He$ \emph{atoms} in the source
($S$) and the detector ($D$), while the internal line describes the propagating
antineutrino. Since we are mainly interested in the phenomenology of \MB\
neutrino oscillations, we avoid an explicit treatment of solid state binding
forces and instead assume the external particles to reside in the ground states
of simple harmonic oscillator potentials, with oscillator frequencies of the
order of the Debye temperature $\Theta_D \sim 600\ \mathrm{K}\simeq 0.05$~eV of
the respective crystals~\cite{Raghavan:2006xf, Potzel:2006ad}. It is known from
the theory of the classical photon \MB\ effect~\cite{Lipkin:1973}, that this
model provides qualitatively correct results, even though it is, of course,
insufficient for computing a precise prediction of the total event rate.  If we
denote the masses of the external particles by $m_A$ ($A = \{ \H, \He \}$),
their average positions by $\vec{x}_B$ ($B = \{ S, D \}$), the harmonic
oscillator frequencies by $\omega_{A,B}$, and the ground state energies by
$E_{A,B}$, the wave functions corresponding to the external legs in
fig.~\ref{fig:feyn} are given by,
\begin{align}
  \psi_{A,B,0}(\vec{x}, t) = \bigg[\frac{m_A \omega_{A,B}}{\pi}\bigg]^\frac{3}{4}
  \exp\bigg[\! -\frac{1}{2} m_A \omega_{A,B} |\vec{x} - \vec{x}_B|^2 \bigg] \, e^{-i E_{A,B} t} \,.
  \label{eq:HO-WF-gs}
\end{align}
Due to interactions of the atoms with their surroundings, $E_{A,B}$ will not be
constant in time, but will fluctuate around the average zero point energy
$E_{A,B,0} = m_A + \frac{1}{2} \omega_{A,B}$~\cite{Coussement:1992a,Coussement:1992b,
Odeurs:1995,Balko:1997,Odeurs:1997}. These fluctuations are, for example,
induced by random thermal spin flips of neighboring atoms. They are generally
referred to as homogeneous line broadening effects because, as we will see
below, they limit the achievable sharpness of the \MB\ resonance.  To describe
homogeneous line broadening, we make the replacement~\cite{Odeurs:1995}
\begin{align}
  e^{-i E_{A,B} t} \rightarrow e^{-i E_{A,B,0} t} \, f_{A,B}(t) \,,
\end{align}
in eq.~\eqref{eq:HO-WF-gs}, where
\begin{align}
  f_{A,B}(t) = \exp\bigg[
                     -i \int_0^t \! dt^\prime \,
                     \big( E_{A,B}(t^\prime) - E_{A,B,0} \big)
                   \bigg] \,
  \label{eq:mod-factors}
\end{align}
is integrated phase shift induced by the fluctuations of $E_{A,B}$.  Note that
this approach accounts only for homogeneous line broadening due to solid-state
effects, but not for broadening due the natural line width.  The latter effect
(which is theoretically interesting, but completely negligible in the
$^3\H$\,--\,$^3\He$ system) has been studied in detail in
ref.~\cite{Akhmedov:2008jn}.  The transition amplitude corresponding to
fig.~\ref{fig:feyn}, including the modulation factors \eqref{eq:mod-factors},
is
\begin{align}
  i \mathcal{A} &=
      \int\! d^3x_1 \, dt_1 \int\! d^3 x_2 \, dt_2 \,
      \bigg( \frac{m_\H  \omega_{\H, S}}{\pi} \bigg)^{\frac{3}{4}}
      \exp\bigg[\! -\frac{1}{2} m_\H \omega_{\H,S}
      |\vec{x}_1 - \vec{x}_S|^2 \bigg] \, f_{\H,S}(t_1) \, e^{-i E_{\H,S} t_1}  \nonumber\\
  &\hspace{1cm} \cdot
      \bigg( \frac{m_\He \omega_{\He,S}}{\pi} \bigg)^{\frac{3}{4}}
      \exp\bigg[\! -\frac{1}{2} m_\He \omega_{\He,S}
      |\vec{x}_1 - \vec{x}_S|^2 \bigg] \, f_{\He,S}^*(t_1) \, e^{+i E_{\He,S} t_1} \nonumber\\
  &\hspace{1cm} \cdot
      \bigg( \frac{m_\He \omega_{\He, D}}{\pi} \bigg)^{\frac{3}{4}}
      \exp\bigg[\! -\frac{1}{2} m_\He \omega_{\He,D}
      |\vec{x}_2 - \vec{x}_D|^2 \bigg] \, f_{\He,D}(t_2) \, e^{-i E_{\He,D} t_2} \nonumber\\
  &\hspace{1cm} \cdot
      \bigg( \frac{m_\H \omega_{\H,D}}{\pi} \bigg)^{\frac{3}{4}}
      \exp\bigg[\! -\frac{1}{2} m_\H \omega_{\H,D}
      |\vec{x}_2 - \vec{x}_D|^2 \bigg] \, f_{\H,D}^*(t_2) \, e^{+i E_{\H,D} t_2}  \nonumber\\
  &\hspace{1cm} \cdot \sum_j
      \mathcal{M}_S^\mu \mathcal{M}_D^{\nu *} \, |U_{ej}|^2 \int \! 
      \frac{d^4p}{(2\pi)^4}
      \exp\big[\! -i p_0 (t_2 - t_1) + i \vec{p} (\vec{x}_2 - \vec{x}_1) \big] \nonumber\\
  &\hspace{1cm} \cdot
      \bar{u}_{e,S} \gamma_\mu (1 - \gamma^5) \,
      \frac{i (\slashed{p} + m_j)}{p_0^2 - \vec{p}^2 - m_j^2 + i\epsilon} \,
      (1 + \gamma^5) \gamma_\nu u_{e,D} \,.
  \label{eq:qft-A1}
\end{align}
Here, $m_j$ are the neutrino mass eigenvalues, $U_{ej}$ are elements of the
leptonic mixing matrix, and the nonrelativistic (i.e.\ momentum-independent)
electron spinors are denoted by $u_{e,S}$ for the electron that is emitted in
$^3\H$ decay in the source, and by $u_{e,D}$ for the electron that is destroyed
in the neutrino capture process in the detector. The matrix elements
$\mathcal{M}^\mu_S$ and $\mathcal{M}^\mu_D$ are given by
\begin{align}
  \mathcal{M}_{S,D}^\mu
    &= \frac{G_F \cos\theta_c}{\sqrt{2}} \,
       \psi_e(R) \, \bar{u}_\He (M_V\, \delta^\mu_0 - g_A M_A \gamma^i \gamma^5 \,
       \delta^{\mu}_i/\sqrt{3} ) u_\H\,\kappa_{S,D}^{1/2} \,,
  \label{eq:qft-MSD}
\end{align}
where $G_F$ is the Fermi constant, $\theta_c$ the Cabibbo angle, and $u_{A,B}$
(with $A = \{ \H, \He\}$, $B = \{ S, D \}$ as before) are the non-relativistic
$^3\H$ and $^3\He$ spinors. The vector and axial vector (or Fermi and
Gamow-Teller) nuclear matrix elements are $M_V = 1$ and $M_A\approx \sqrt{3}$,
respectively~\cite{Perkins:HEP3, Povh:ParticlesAndNuclei},
and the axial-vector coupling constant is $g_A \simeq 1.25$.  The quantity
$\psi_e(R)$ gives the value of the anti-symmetrized atomic wave function of
$^3\He$ at the surface of the nucleus, while the factor
\begin{align}
  \kappa_{S,D} &= \bigg| \int \Psi_{Z=2,S,D}(\vec{r})^* \,
                          \Psi_{Z=1,S,D}(\vec{r})\, d^3 r \,\bigg|^2\,.
  \label{eq:qft-kappaS}
\end{align}
accounts for the fact that the spectator electron in bound state $^3\H$ decay
and induced orbital electron capture on $^3\He$ changes from the $1s$ state of
$^3\H$ into the $1s$ state of $^3\He$, or vice-versa.

The spatial integrals in \eqref{eq:qft-A1} yield a factor
$\exp[-\vec{p}^2/2\sigma_p^2] \exp[i \vec{p} \vec{L}]$, with the effective
momentum uncertainty $\sigma_p$ of the experiment defined by
\begin{align}
  \frac{1}{\sigma_p^2}
    &= \frac{1}{\sigma_{pS}^2} + \frac{1}{\sigma_{pD}^2}
     =   \frac{1}{m_\H \omega_{\H,S} + m_\He \omega_{\He,S}}
       + \frac{1}{m_\H \omega_{\H,D} + m_\He \omega_{\He,D}}\,,
  \label{eq:sigma-p}
\end{align}
and with the baseline vector
\begin{align}
  \vec{L} = \vec{x}_D - \vec{x}_S \,.
\end{align}
To evaluate the three-momentum integral, we employ the Grimus-Stockinger
theorem~\cite{Grimus:1996av}, which states that, for any three times
continuously differentiable function $\psi(\vec{p})$ ($\vec{p} \in \mathbb{R}^3$),
with $\psi$ and all its first and second derivatives decreasing at least as
$1/|\vec{p}|^2$ for $|\vec{p}| \rightarrow \infty$, the following relation
holds for any real number $A > 0$:
\begin{align}
  \int d^3p \, \frac{\psi(\vec{p}) \, e^{i \vec{p} \vec{L}}}{A - \vec{p}^2 + i\epsilon}
    \xrightarrow{|\vec{L}| \rightarrow \infty}
    -\frac{2 \pi^2}{L} \psi(\sqrt{A} \tfrac{\vec{L}}{L}) e^{i \sqrt{A} L}
    + \mathcal{O} (L^{-\frac{3}{2}}) \,.
  \label{eq:Grimus}
\end{align}
Effectively, this formula gives the form of the Feynman propagator for
propagation over macroscopic distances.  The parameter $A$ corresponds to
squared modulus of the on-shell momentum component of the propagating particle.
Applying the Grimus-Stockinger theorem to our expression for $\mathcal{A}$, we
find
\begin{align}
  i \mathcal{A} &=
      \frac{-i}{8\pi^2 L} \mathcal{N}
      \sum_j \mathcal{M}_S^\mu \mathcal{M}_D^{\nu *} |U_{ej}|^2 \!
      \int_{-\infty}^\infty \! dt_1 \, dt_2 \,
      f_{\H,S}(t_1) \, f_{\He,S}^*(t_1) \, f_{\He,D}(t_2) \, f_{\H,D}^*(t_2)
                                                  \nonumber\\
  &\quad \cdot
      \int_{-\infty}^\infty \!\! dp_0 \,
      \exp\bigg[\! -\frac{p_0^2 - m_j^2}{2 \sigma_p^2} \bigg] \, 
       e^{i \sqrt{p_0^2 - m_j^2} L}
       e^{-i (E_{S,0} - p_0) t_1 + i (E_{D,0} - p_0) t_2}
                                                  \nonumber\\[0.3cm]
  &\quad \cdot
      \bar{u}_{e,S} \gamma_\mu (1 - \gamma^5) (\slashed{p}_j + m_j)
        (1 + \gamma^5) \gamma_\nu u_{e,D} \,,
  \label{eq:qft-A2}
\end{align}
with $p_j \equiv (p_0, \sqrt{p_0^2 - m_j^2} \, \vec{L} / L)$ and with the
constant
\begin{align}
  \mathcal{N} &=
    \bigg( \frac{m_\H  \omega_{\H, S}}{\pi} \bigg)^{\frac{3}{4}}
    \bigg( \frac{m_\He \omega_{\He,S}}{\pi} \bigg)^{\frac{3}{4}}
    \bigg( \frac{m_\He \omega_{\He,D}}{\pi} \bigg)^{\frac{3}{4}}
    \bigg( \frac{m_\H  \omega_{\H, D}}{\pi} \bigg)^{\frac{3}{4}} \nonumber\\
  &\hspace{3cm} \cdot
    \bigg( \frac{2\pi}{m_\H \omega_{\H,S} + m_\He \omega_{\He,S}} 
                                            \bigg)^\frac{3}{2}
    \bigg( \frac{2\pi}{m_\H \omega_{\H,D} + m_\He \omega_{\He,D}} 
                                    \bigg)^\frac{3}{2} \,
\end{align}
containing the wavefunction normalization factors from eq.~\eqref{eq:HO-WF-gs}
and the numerical prefactors that have arisen in the $\vec{x}_1$ and $\vec{x}_2$
integrations.  Since we do not know the exact form of the modulation factors
$f_{A,B}(t)$, we cannot evaluate the time integrals at this stage. However,
ultimately, we are only interested in the transition rate $\Gamma$, which is
proportional to $\ev{\mathcal{A} \mathcal{A}^*}$, the statistical average of
$\mathcal{A} \mathcal{A}^*$ over all possible $^3\H$ and $^3\He$ states in the
source and the detector. This expression can be simplified using statistical
arguments.  In particular, when evaluating it, we encounter the quantity
\begin{align}
  B_S(t_1, \tilde{t}_1) \equiv
  \Big\langle
    f_{\H,S}(t_1) \, f_{\He,S}^*(t_1) \, f_{\H,S}^*(\tilde{t}_1) \, f_{\He,S}(\tilde{t}_1) \,
  \Big\rangle
  = \bigg\langle \! \exp\bigg[
      -i \int_{\tilde{t}_1}^{t_1} \! dt^\prime \, \Delta E_S(t^\prime)
    \bigg] \bigg\rangle \,,
  \label{eq:qft-avg}
\end{align}
and a similar term from the detector-related modulation factors.  Here, $t_1$
and $\tilde{t}_1$ are the time variables appearing in the expressions for
$\mathcal{A}$ and $\mathcal{A}^*$, respectively. To shorten the notation, we
have defined a quantity $\Delta E_S(t^\prime) \equiv E_S(t^\prime) - E_{S,0} \equiv
[E_{\H,S}(t^\prime) - E_{\He,S}(t^\prime)] - [E_{\H,S,0} - E_{\He,S,0}]$, which
gives the deviation of the energy of the neutrino emission line from its mean
value at time $t^\prime$.  Following \cite{Odeurs:1995}, we assume $\Delta
E_S(t^\prime)$ to be a Gaussian random variable centered around zero:
\begin{align}
  \ev{\Delta E_S(t^\prime)} = 0 \,.
  \label{eq:qft-1point}
\end{align}
Moreover, we assume fluctuations at different points in time to be uncorrelated
(Markov\-ian approximation), which implies
\begin{align}
  \Big\langle \Delta E_S(t^\prime) \, \Delta E_S(t^{\prime\prime}) \Big\rangle
    = \gamma_S \, \delta(t^\prime - t^{\prime\prime}) \,.
  \label{eq:qft-2point}
\end{align}
This is a good approximation if the correlation time of the fluctuations is
much smaller than all other time scales appearing in the problem, in particular
the tritium life time and the running time of the experiment.  The constant
$\gamma_S$ will turn out to be the width of the neutrino emission line.
Proceeding along the lines of refs.~\cite{Odeurs:1995, Meystre:QO}, we expand
\eqref{eq:qft-avg} into a Taylor series and obtain
\begin{align}
  B_S(t_1, \tilde{t}_1) =
    \sum_{n=0}^{\infty} \, \frac{(-i)^n}{n!} \,
    \int_{\tilde{t}_1}^{t_1} \! dt^{(1)} \cdots dt^{(n)}
    \ev{\, \Delta E_S(t^{(1)}) \cdots \Delta E_S(t^{(n)})} \,.
\end{align}
One can now use the assumption that $\Delta E_S(t^{(i)})$ is normally
distributed around zero to show that the $n$-point correlation functions on the
right hand side can, for even $n$, be rewritten by splitting them into products
of two-point functions (which can be evaluated by using
\eqref{eq:qft-2point}) and summing over all $(n-1)(n-3) \cdots 3 \cdot 1 =
n! / [2^{n/2} (n/2)!]$ distinct combinations of such two-point functions.
For odd $n$, the $n$-point correlation functions can be transformed into
products of $(n-1)/2$ two-point functions and a one-point function, which is
zero by virtue of eq.~\eqref{eq:qft-1point}.  Therefore, $B_S(t_1,
\tilde{t}_1)$ takes the form
\begin{align}
  B_S(t_1, \tilde{t}_1)
    &= \sum_{n=0}^{\infty} \, \frac{(-\gamma_S / 2)^n}{n!} \,
       \prod_{i=1}^{n} \, \int_{\tilde{t}_1}^{t_1} \!
       dt^{(2i)} \, dt^{(2i - 1)} \,
       \delta(t^{(2i)} - t^{(2i-1)})                         \nonumber\\
    &= \exp \Big[ -\frac{1}{2} \gamma_S |t_1 - \tilde{t}_1| \Big] \,.
\end{align}
The analogous expression for the detector-related modulation factors is
\begin{align}
  B_D(t_2, \tilde{t}_2)
    &= \exp \Big[ -\frac{1}{2} \gamma_D |t_2 - \tilde{t}_2| \Big] \,.
\end{align}
Using
\begin{align}
  \int_{-\infty}^\infty \!\! dt_1 \, d\tilde{t}_1 \, dt_2 \, d\tilde{t}_2
    &\exp\Big[-\frac{1}{2} \gamma_S |t_1 - \tilde{t}_1| - i (E_{S,0} - p_0) t_1
              + (E_{S,0} - \tilde{p}_0) t_1 \Big] \nonumber\\
    \cdot&
     \exp\Big[-\frac{1}{2} \gamma_D |t_2 - \tilde{t}_2| + i (E_{D,0} - p_0) t_2
              - (E_{D,0} - \tilde{p}_0) t_2 \Big] \nonumber\\
  &\hspace{-1cm} =
    (2\pi)^4 [\delta(p_0 - \tilde{p}_0)]^2
    \frac{\gamma_S / 2\pi}{(E_{S,0} - p_0)^2 + \gamma_S^2/4}
    \frac{\gamma_D / 2\pi}{(E_{D,0} - p_0)^2 + \gamma_D^2/4} \,,
\end{align}
the expression for $\ev{\mathcal{A} \mathcal{A}^*}$ now becomes
\begin{align}
  \ev{\mathcal{A} \mathcal{A}^*}
  &= \frac{\mathcal{N}^2}{64 \pi^4 L^2}
     \sum_{j,k} \mathcal{M}_S^\mu \mathcal{M}_D^{\nu *}
                \mathcal{M}_S^{\rho*} \mathcal{M}_D^{\sigma}
     |U_{ej}|^2 |U_{ek}|^2 \!
     \int_{-\infty}^\infty \!\! dp_0 \, d\tilde{p}_0
     \exp\bigg[\! -\frac{2p_0^2 - m_j^2 - m_k^2}{2 \sigma_p^2} \bigg] \, 
                                                  \nonumber\\
  &\cdot
     (2\pi)^4 [\delta(p_0 - \tilde{p}_0)]^2
     \frac{\gamma_S / 2\pi}{(E_{S,0} - p_0)^2 + \gamma_S^2/4}
     \frac{\gamma_D / 2\pi}{(E_{D,0} - p_0)^2 + \gamma_D^2/4}
     e^{i \big( \sqrt{p_0^2 - m_j^2} - \sqrt{p_0^2 - m_k^2} \big) \!L}
                                                  \nonumber\\
  &\cdot
     \bar{u}_{e,S} \gamma_\mu (1 - \gamma^5) (\slashed{p}_j + m_j)
        (1 + \gamma^5) \gamma_\nu u_{e,D}        
     \bar{u}_{e,D} \gamma_\sigma (1 - \gamma^5) (\slashed{\tilde{p}}_j + m_k)
        (1 + \gamma^5) \gamma_\rho u_{e,D} \,.
  \label{eq:qft-tInt}
\end{align}
We can rewrite the squared $\delta$-function as $T/2\pi \cdot \delta(p_0 -
\tilde{p}_0)$ (with $T$ the total running time of the experiment), and use the
remaining $\delta$-factor to evaluate the $\tilde{p}_0$ integral.  We are left
with the $p_0$ integration, which receives its main contribution from the
region where $|E_{S,0} - p_0| \lesssim \gamma_S$ and $|E_{D,0} - p_0| \lesssim
\gamma_D$ due to the Lorentzians on the right hand side of
eq.~\eqref{eq:qft-tInt}.  Since $\gamma_{S,D} \ll \sigma_p$ and $\gamma_{S,D}
\ll E_{S,0}, E_{D,0}$, the spinorial factors as well as the real exponential that will
lead to the generalized Lamb-\MB\ factor and to the localization term are
almost constant over this region and may be replaced by their values at
\begin{align}
  \bar{E} = \frac{1}{2} (E_{S,0} + E_{D,0}) \,.
  \label{eq:Ebar}
\end{align}
If we finally expand the oscillation phase in $\Delta m_{jk}^2/p_0^2$, the
$p_0$ integral becomes~\cite{Akhmedov:2008jn}
\begin{align}
  I_{jk} &\equiv
    \int_{-\infty}^\infty \! dp^0 \, 
    \frac{\gamma_S/2\pi}{(p^0 - E_{S,0})^2 + \gamma_S^2/4} \,
    \frac{\gamma_D/2\pi}{(p^0 - E_{D,0})^2 + \gamma_D^2/4} \,
    \exp\bigg[\! -i \frac{\Delta m_{jk}^2 L}{2 p^0} \bigg] \,  \nonumber\\
  &= \frac{1}{2\pi}
    \frac{1}{(E_{S,0} - E_{D,0})^2 + \tfrac{(\gamma_S + \gamma_D)^2}{4}} \,
    \Bigg\{
      \frac{\gamma_S + \gamma_D}{2} (A_{jk}^{(S)} + A_{jk}^{(D)}) \nonumber\\
  &\hspace{2cm}
    - \frac{1}{2} \frac{(A_{jk}^{(S)} - A_{jk}^{(D)})
                        \big[
                          (E_{S,0} - E_{D,0})(\gamma_S - \gamma_D)
                          \pm i \frac{(\gamma_S + \gamma_D)^2}{2}
                        \big]}
                       {E_{S,0} - E_{D,0} \pm i \, \tfrac{\gamma_S - \gamma_D}{2}}
    \Bigg\} \,,
  \label{eq:qft-Ijk1}
\end{align}
with the abbreviations
\begin{align}
  A_{jk}^{(B)}
    &= \exp\bigg[ -i \frac{\Delta m_{jk}^2}{2(E_{B,0} \pm i \, 
       \tfrac{\gamma_B}{2})}\, L \bigg]        \nonumber\\
    &\simeq \exp\bigg[ -2\pi i \frac{L}{L^{\rm osc}_{B,jk}} \bigg] \, 
            \exp\bigg[ - \frac{L}{L^{\rm coh}_{B,jk}} \bigg] \,,
  \label{eq:qft-abbrev-A}
\end{align}
and with the oscillation and coherence lengths
\begin{align}
  L^{\rm osc}_{B,jk} = \frac{4\pi E_{B,0}}{\Delta m_{jk}^2}
                       \simeq  \frac{4\pi \bar{E}}{\Delta m_{jk}^2}\,
  \qquad\text{and}\qquad
  L^{\rm coh}_{B,jk} = \frac{4 E_{B,0}^2}{\gamma_B |\Delta m_{jk}^2|}
                       \simeq \frac{4 \bar{E}^2}{\gamma_B |\Delta m_{jk}^2|} \,.
  \label{eq:qft-inh-Lcoh}
\end{align}
In eq.~\eqref{eq:qft-Ijk1}, the upper (lower) signs correspond to $\Delta
m_{jk}^2 > 0$ ($\Delta m_{jk}^2 < 0$).  Thus, the transition rate $\Gamma$ for
a \MB\ neutrino experiment dominated by homogeneous line broadening is, according
to Fermi's Golden Rule,
\begin{align}
   \Gamma &=
     \frac{\Gamma_0 \,B_0}{4\pi L^2}\;Y_S Y_D \,\frac{1}{2\pi}\,
       \sum_{j,k} |U_{ej}|^2 |U_{ek}|^2 \,
       \exp\bigg[\! -\frac{2 \bar{E}^2 - m_j^2 - m_k^2}{2 \sigma_p^2} \bigg]
       \frac{1}{(E_{S,0} - E_{D,0})^2 + \tfrac{(\gamma_S + \gamma_D)^2}{4}}
                                                           \nonumber\\
    &\cdot
       \Bigg[
         \frac{\gamma_S + \gamma_D}{2} (A_{jk}^{(S)} + A_{jk}^{(D)})
       - \frac{1}{2} \frac{(A_{jk}^{(S)} - A_{jk}^{(D)})
                           \big[
                             (E_{S,0} - E_{D,0})(\gamma_S - \gamma_D)
                             \pm i \frac{(\gamma_S + \gamma_D)^2}{2}
                           \big]}
                          {E_{S,0} - E_{D,0} \pm i \, \tfrac{\gamma_S - \gamma_D}{2}}
       \Bigg] \,,
  \label{eq:qft-Gamma}
\end{align}
where
\begin{align}
  \Gamma_0 &\equiv \frac{G_F^2 \cos^2\theta_c}{\pi} \; |\psi_e(R)|^2 \,m_e^2\, 
              \big( |M_V|^2+g_A^2 |M_A|^2 \big) \,
              \bigg( \frac{E_{S,0}}{m_e} \bigg)^2 \kappa_S \,
  \label{eq:qft-Gamma0}
\end{align}
is the rate of bound state $^3\H$ decay, and
\begin{align}
  B_0 &\equiv 4\pi G_F^2 \cos^2\theta_c\; |\psi_e(R)|^2  
         \big( |M_V|^2 + g_A^2 |M_A|^2 \big) \, \kappa_D \,
  \label{eq:qft-B0}
\end{align}
is related to the cross section for induced orbital electron capture
on \emph{free} $^3\He$ by~\cite{Mikaelyan:1967}
\begin{align}
  \sigma(E_\nu) = B_0 \, \rho(E_{\bar{\nu}, {\rm res}}) \,.
\end{align}
Here $\rho(E_{\bar{\nu}, {\rm res}})$ is the spectral density of incident
neutrinos, i.e.\ the number of neutrinos per unit energy interval, at the
resonance energy for this case, $E_{\bar{\nu}, {\rm res}} = Q + E_R$ (where $Q
= 18.6$~keV is the $Q$-value of the process and $E_R$ is the recoil energy
transferred to the atom).  The quantities $Y_S$ and $Y_D$ in
eq.~\eqref{eq:qft-Gamma} are given by
\begin{align}
  Y_B = 8 \bigg( \sqrt{\frac{m_\H \,\omega_{\H,B}}{m_\He\, \omega_{\He,B}}}
              +  \sqrt{\frac{m_\He \,\omega_{\He,B}}{m_\H\, \omega_{\H,B}}} \bigg)^{-3} \,
\end{align}
for $B = \{ S, D \}$.

As anticipated, \eqref{eq:qft-Gamma} coincides precisely with the
corresponding expression for the case of inhomogeneous line broadening, given
in eq.~(41) of ref.~\cite{Akhmedov:2008jn}.%
\footnote{In the present work, we have chosen to present $\Gamma$ in a form where
the Breit-Wigner term is factorized out of the term containing the oscillation
and coherence exponentials. It is straightforward to check that this form is
identical to the form used in ref.~\cite{Akhmedov:2008jn}.}
We find again a Breit-Wigner-like
resonance term, which suppresses \MB\ transitions if the central energies
$E_{S,0}$ and $E_{D,0}$ of the emission and absorption lines differ
by more than the average line width $(\gamma_S + \gamma_D)/2$, and a factor
\begin{align}
  \exp\bigg[\! -\frac{2 \bar{E}^2 - m_j^2 - m_k^2}{2 \sigma_p^2} \bigg]
    &= \exp\bigg[ -\frac{(p^{\rm min}_{jk})^2}{\sigma_p^2} \bigg]
       \exp\bigg[ -\frac{|\Delta m_{jk}^2|}{2 \sigma_p^2} \bigg] \,,
  \label{eq:qft-Lamb-MB}
\end{align}
with
\begin{align}
  (p^{\rm min}_{jk})^2 = \bar{E}^2 - \max(m_j^2, m_k^2)\,,
\end{align}
which we interpret as a generalized Lamb-\MB\ factor (or fraction of
recoil-free emissions/absorptions), multiplied with a localization term. The
latter can be neglected if $\sigma_p^2 \gg \Delta m_{jk}^2$, or equivalently,
if $L^{\rm osc}_{B,jk} \gg 4\pi \sigma_x \bar{E} / \sigma_p$, where $\sigma_x
\equiv 1/2 \sigma_p$ is the spatial delocalization of the emitting and
absorbing atoms. For $\bar{E} = 18.6$~keV and $\sigma_p \sim (m_{H}
\theta_D)^{1/2} \sim 7$~keV, it is clear that this inequality is easily
fulfilled since $\sigma_x$ is of the order of the interatomic distance, while
$L^{\rm osc}_{B,jk} \sim 20$~m for oscillations driven by the atmospheric mass
squared difference $\Delta m_{31}^2$ and $L^{\rm osc}_{B,jk} \sim 600$~m for
oscillations driven by the solar mass squared difference $\Delta m_{21}^2$.
The factors $A_{jk}^{(B)}$ in eq.~\eqref{eq:qft-Gamma} contain the oscillation
exponentials and the decoherence terms which describe the effect of wave
packet separation due to the different group velocities associated with
different neutrino mass eigenstates. However, it is easy to see that
decoherence is not an issue in any realistic \MB\ neutrino experiment because
the corresponding coherence lengths are of $\mathcal{O}(10^{13}\ {\rm km})$.

The fact that the formula for $\Gamma$ is identical for the cases of
homogeneous and inhomogeneous line broadening implies that these two situations
cannot be distinguished experimentally. This confirms a more general theorem by
Kiers, Nussinov, and Weiss~\cite{Kiers:1995zj} which states that it is
impossible to distinguish an ensemble of neutrino wave packets with identical
momentum distributions from an ensemble of plane wave neutrinos whose
individual momenta follow the same distribution.  In fact, the density matrix
describing the ensemble is identical for both cases.  Applied to \MB\
neutrinos, the case of neutrino wave packets corresponds to a situation where
homogeneous line broadening is dominant, so that each neutrino wave packet is
broadened because the energy of the emission line, $E_S$, changes during the
emission process. In contrast, for mostly inhomogeneous line broadening, each
individual neutrino can be approximately described by a plane wave because it
is emitted with an extremely small energy spread (which is ultimately
determined by subdominant homogeneous solid state effects, by the natural
width, and by the Heisenberg principle). Different neutrinos, however, are
emitted with different energies which depend, for example, on the proximity of
the emitting atom to crystal impurities and lattice defects.

To end this section, let us give a simpler and more useful form of
eq.~\eqref{eq:qft-Gamma}, obtained by neglecting the localization and coherence
terms and considering the two-flavor approximation, with an effective mixing
angle $\theta$, an effective mass squared difference $\Delta m^2$, and an
average absolute neutrino mass $\bar{m}$~\cite{Akhmedov:2008jn}:
\begin{align}
  \Gamma &\simeq 
    \frac{\Gamma_0 \,B_0}{4\pi L^2} \; Y_S Y_D \,
    \exp\bigg[\! -\frac{\bar{E}^2 - \bar{m}^2}{\sigma_p^2} \bigg] \,
    \frac{(\gamma_S + \gamma_D) / 2\pi}
         {(E_{S,0} - E_{D,0})^2 + \frac{(\gamma_S + \gamma_D)^2}{4}} 
    \bigg\{
      1 - \sin^2 2\theta \, \sin^2 \bigg( \pi \frac{L}{L^{\rm osc}} \bigg)
    \bigg\} \,.
  \label{eq:qft-Gamma-2f}
\end{align}

\section{\MB\ neutrinos in quantum mechanics: Lorentzian wave packets}
\label{sec:qm}

Let us now discuss how oscillations of \MB\ neutrinos can be understood in the
framework of quantum mechanics. Since QM is unable to describe particle
creation and destruction, we cannot directly include the production and
detection processes into our formalism, as in QFT. Instead, we will first
compute the probability for transitions between the initial and final neutrino
states, and then multiply this with the emitted flux and with the absorption
cross section to obtain the overall event rate $\Gamma$.  We will describe the
propagating neutrino as a superposition of three wave packets, one for each
mass eigenstate~\cite{Giunti:1991ca, Giunti:1991sx, Kiers:1995zj,
Giunti:1997wq, Giunti:2002ee, Giunti:2003ax}. As we have discussed above, such
a description corresponds to \MB\ neutrinos in the regime of homogeneous line
broadening, while the case of inhomogeneous broadening would be more naturally
implemented by considering an ensemble of many plane wave neutrinos in the
density matrix approach~\cite{Kiers:1995zj}.  However, since homogeneous and
inhomogeneous line broadening cannot be distinguished experimentally, it is
sufficient to focus on one of the two cases. We use the wave packet picture
because it provides insights into the evolution of each single neutrino, which
we find useful to better understand the localization and coherence conditions
that will emerge.

Unlike most other authors, who use wave packets with a Gaussian shape, we will
use wave packets with a Lorentzian momentum distribution because it is known
from the classical \MB\ effect that homogeneous and inhomogeneous line
broadening mechanisms lead to a Lorentzian energy
spread~\cite{Frauenfelder:1962, Potzel:PrivComm}. The momentum space wave
function for the electron antineutrino produced in $^3\H$ decay thus has the
form
\begin{align}
  \sprod{p}{\bar{\nu}_{e S}(t)}
    &= \frac{1}{N_S} \, \sum_{j} U_{ej} \, f_{jS} \,
       \frac{\sqrt{\gamma_S/2\pi}}{p - p_{jS} + i \gamma_S /2} \,
       \exp\big[- i E_j t \big] \,
       \ket{\nu_j} \,.
  \label{eq:qm-Lorentz-1}
\end{align}
The index $S$ indicates that this state is produced in the neutrino source, and
the normalization factor is $N_S = \big(\sum_{j} |U_{ej}|^2 \,
|f_{jS}|^2\big)^{1/2}$.  Similarly, the detection process can be described as a
projection of $\ket{\bar{\nu}_{eS}(t)}$ onto a state $\ket{\bar{\nu}_{eD}}$
with the momentum space representation
\begin{align}
  \sprod{p}{\bar{\nu}_{eD}}
    &= \frac{1}{N_D} \, \sum_{j} U_{ej} \, f_{jD} \,
       \frac{\sqrt{\gamma_D/2\pi}}{p - p_{jD} + i \gamma_D /2} \,
       \exp\big[- i p L \big] \,
       \ket{\nu_j} \,
  \label{eq:qm-Lorentz-3}
\end{align}
and the normalization factor $N_D = \big(\sum_{j} |U_{ej}|^2 \,
|f_{jS}|^2\big)^{1/2}$.  In the above expressions, $p_{jS}$, $p_{jD}$ are the
central momenta of the wave packets, $E_j = (p^2 + m_j^2)^{1/2}$, and
$\gamma_S$, $\gamma_D$ are the wave packet widths. Moreover, we have introduced
phenomenological fudge factors $f_{jS}$, $f_{jD}$ that will be motivated and
discussed below.

For \MB\ neutrinos, $\gamma_S$ and $\gamma_D$ are of the order of the energy
uncertainty associated with the emission and detection processes, which is of
order $10^{-11}$~eV. The much larger \emph{momentum} uncertainties of the
source and the detector do not play a role because the neutrino is on-shell,
so that by virtue of the relativistic energy-momentum relation the momentum
uncertainty of the neutrino cannot be larger than its energy uncertainty. (Of
course, the momenta associated with the different mass eigenstates have to
differ by much more than $10^{-11}$~eV in order to ensure energy-momentum
conservation in the production and detection processes.)

Note that $\ket{\bar{\nu}_{eD}}$ is time-independent (on this point, we disagree
with ref.~\cite{Giunti:2003ax}, where the detection operator $\ket{\nu_{\beta D}}
\bra{\nu_{\beta D}}$ is assumed to be not a time-independent but only a
time-averaged quantity); on the other hand, a factor $\exp[- i p L]$ is
required to center the wave packet around $x = L$.

The phenomenological fudge factors $f_{jS}$ and $f_{jD}$ can be used to
describe a possible mass dependence of the neutrino production and detection
amplitudes.  For example, we have seen in the previous section that the
Lamb-\MB\ factor depends on $m_j$, so that the production and absorption of the
lighter neutrino mass eigenstates is slightly suppressed compared to the
production and absorption of the heavier ones. This can be viewed as a slight
dynamical reduction of neutrino mixing. Let us stress that $f_{jS}$ and
$f_{jD}$ cannot be determined in the QM approach, and have to be put in by
hand. We will choose
\begin{align}
  f_{jS} \equiv \exp\bigg[\! \frac{\bar{E}^2 - m_j^2}{2 \sigma_{pS}^2} \bigg] \,,
  \qquad\qquad
  f_{jD} \equiv \exp\bigg[\! \frac{\bar{E}^2 - m_j^2}{2 \sigma_{pD}^2} \bigg] \,
\end{align}
(with $\sigma_{pS}$, $\sigma_{pD}$, and $\bar{E}$ defined as in
eqs.~\eqref{eq:sigma-p} and \eqref{eq:Ebar}, respectively) in order to
ultimately reproduce the correct Lamb-\MB\ factor.

The amplitude for the transition $\ket{\bar{\nu}_{eS}(t)} \rightarrow
\ket{\bar{\nu}_{eD}}$ is given by
\begin{align}
  \mathcal{A}(t, L) = \sprod{\bar{\nu}_{eD}}{\bar{\nu}_{eS}(t)} =
  \int\! dp \, \sprod{\bar{\nu}_{eD}}{p} \sprod{p}{\bar{\nu}_{eS}(t)} \,.
  \label{eq:qm-A1}
\end{align}
To be able to evaluate this integral, we make use of the smallness of $\gamma_S$
and $\gamma_D$, and expand $E_j$ around the average momentum $\bar{p}_j = (p_{jS} +
p_{jD})/2$, which gives
\begin{align}
  E_j = \sqrt{p^2 + m_j^2} \; t  \simeq
    \bar{E}_{j} \, t  +  \bar{v}_j t (p - \bar{p}_j) \,,
  \label{eq:qm-Ej-expansion}
\end{align}
with the definitions
\begin{align}
  \bar{E}_j = \sqrt{\bar{p}_j^2 + m_j^2} \qquad\text{and}\qquad
  \bar{v}_j = \frac{\bar{p}_j}{\sqrt{\bar{p}_j^2 + m_j^2}} \,.
  \label{eq:EjS-vjS}
\end{align}
This approximation corresponds to neglecting dispersion (wave packet
spreading), which is a second-order effect~\cite{Beuthe:2001rc}.
Eq.~\eqref{eq:qm-Ej-expansion} is a good approximation as long as $(p -
\bar{p}_j)/\bar{E}_j \ll \bar{E}_j^2 / m_j^2$ for all $p$ within the peak
regions of the source and detector wave packets. We can now compute
$\mathcal{A}(t, L)$, and obtain
\begin{align}
  \mathcal{A}(t, L)
  &= \frac{1}{N_S N_D}
     \sum_{j} |U_{ej}|^2 \, f_{jS} f_{jD}^* \,
     \frac{-i \gamma_S}{p_{jS} - p_{jD} - i (\gamma_S + \gamma_D)/2}
     \exp\big[ - i \bar{E}_j t + i \bar{v}_j \bar{p}_j t \big]
                                                         \nonumber\\
  &\hspace{-0.7cm}\cdot
     \bigg\{\!
       \exp\Big[
             \Big(i p_{jS} + \frac{\gamma_S}{2} \Big)(L - \bar{v}_j t)
           \Big] \theta(\bar{v}_j t - L)
     + \exp\Big[
             \Big( i p_{jD} - \frac{\gamma_D}{2} \Big)(L - \bar{v}_j t)
           \Big] \theta(- \bar{v}_j t + L)
     \bigg\} \,,
  \label{eq:qm-Lorentz-A1}
\end{align}
where $\theta$ denotes the Heaviside step function.

The next step is to compute the transition probability for the process
$\ket{\bar{\nu}_{eS}} \rightarrow \ket{\bar{\nu}_{eD}}$, defined by
\begin{align}
  \mathcal{P}(L) =
    \frac{1}{T} \int_{-T/2}^{T/2}\!dt \,
    \mathcal{A}^*(t,L) \, \mathcal{A}(t,L) \,.
  \label{eq:qm-P-1}
\end{align}
Here, the incoherent averaging over the  running time $T$ of the experiment
reflects the fact that we do not precisely know at which point in time the
production and detection reactions take place.  (The detection time is, of
course, implicitly constrained by the fact that the neutrino wave packet has
sizeable overlap with the detector only during a very short time interval.)
Physically, $\mathcal{P}(L)$ gives the time-averaged probability that a
neutrino prepared in the state $\ket{\bar{\nu}_{e S}(0)}$ at $t=0$ is detected
as $\ket{\bar{\nu}_{e D}}$ at a later time.  Note that $\mathcal{P}(L)$ is
\emph{not} a $\bar{\nu}_e$ survival probability in the usual sense because in
general $\mathcal{P}(L)\,|_{\Delta m_{jk}^2 = 0} \neq 1$. In particular,
irrespective of the neutrino mixing parameters, $\mathcal{P}(L)$ can only be
sizeable if the wave packets $\ket{\bar{\nu}_{e S}(t)}$ and $\ket{\bar{\nu}_{e
D}}$ have sufficient overlap in momentum space.  This is precisely the \MB\
resonance condition.

The experimentally observable event rate $\Gamma$ is obtained by multiplying
$\mathcal{P}(L)$ with the \MB\ neutrino emission rate $\Gamma_0^{\rm MB}$, the
\MB\ neutrino detection cross section $\sigma^{\rm MB}$, and the geometrical
flux suppression factor $1/4\pi L^2$:
\begin{align}
  \Gamma
    &= \frac{1}{4\pi L^2} \, \Gamma_0^{\rm MB} \, \mathcal{P}(L) \, \sigma^{\rm MB}
                                                 \label{eq:fact} \\
    &\equiv
       \frac{1}{4\pi L^2} 
       \Big( \Gamma_0 \, Y_S \, \sum_{j} |U_{ej}|^2 |f_{jS}|^2 \Big) \,
       \mathcal{P}(L) \,
       \Big( B_0 Y_D \frac{T}{2\pi} \, \sum_{j} |U_{ej}|^2 |f_{jD}|^2 \Big) \,.
  \label{eq:qm-Gamma-1}
\end{align}
The parenthesized expressions for $\Gamma_0^{\rm MB}$ and $\sigma^{\rm MB}$
have to be derived in the QFT formalism discussed in sec.~\ref{sec:qft}
and ref.~\cite{Akhmedov:2008jn}. It is impossible to derive them in QM because
they describe particle creation and annihilation processes. Note that we
are here using the cross section for the limiting case of
an infinitely sharp \MB\ resonance --- hence the factor $T/2\pi$, which should
be understood as an approximate $\delta$-peak of the form
\begin{align}
  \delta(0)
    \simeq  \lim_{E \rightarrow E_{D,0}} \, \int_{-T/2}^{T/2} \! dt \, e^{i (E - E_{D,0}) t}
      =     \frac{T}{2\pi} \,.
\end{align}
The effect of line broadening is already accounted for by the fact that
$\mathcal{A}(t,L)$ is suppressed if $|p_{jS} - p_{jD}| \gg (\gamma_S +
\gamma_D)/2$ (cf.~eq.~\eqref{eq:qm-Lorentz-A1}).

Evaluation of $\Gamma$ requires splitting the time integral in
eq.~\eqref{eq:qm-P-1} into three separate integrals with integration domains
$(-\infty, L/\bar{v}_k]$, $(L/\bar{v}_k, L/\bar{v}_j)$, $[L/\bar{v}_j, \infty)$
for $m_j > m_k$, and $(-\infty, L/\bar{v}_j]$, $[L/\bar{v}_j, L/\bar{v}_k]$,
$[L/\bar{v}_k, \infty)$ for $m_j < m_k$. (It is justified to replace the
integration boundaries $\pm T/2$ from eq.~\eqref{eq:qm-P-1} by infinity here
because the overlap of the wave packets $\ket{\bar{\nu}_{e S}(t)}$ and
$\ket{\bar{\nu}_{e D}}$ decreases exponentially at large $T$, when the neutrino
has long passed the detector.) We will only show how to evaluate one of the
above integrals, since the others are similar.  Consider
\begin{align}
  J_{jk} = \int_{L/\bar{v}_k}^{L/\bar{v}_j} \!dt\,
  \exp\bigg[
       &- i (\bar{E}_j - \bar{E}_k) t
        + i (\bar{v}_j \bar{p}_j - \bar{v}_k \bar{p}_k) t
        - i (\bar{v}_j p_{jD} - \bar{v}_k p_{kS}) t
                                                         \nonumber\\
       &+ \frac{1}{2} (\gamma_D \bar{v}_j - \gamma_S \bar{v}_k) t
        + i (p_{jD} - p_{kS}) L
        - \frac{1}{2} (\gamma_D - \gamma_S) L
      \bigg] \,
  \label{eq:Jjk-1}
\end{align}
for $m_j > m_k$.  We use the approximation of ultrarelativistic neutrinos ($m_j
\ll \bar{E}_j$), which suggests the expansions
\begin{align}
  p_{jS}    \simeq E_{S,0} - (1 - \xi_S) \frac{m_j^2}{2E_{S,0}} \,, \qquad\quad
  p_{jD}    \simeq E_{D,0} - (1 - \xi_D) \frac{m_j^2}{2E_{D,0}} \,, \label{eq:qm-rel-pjS}
\end{align}
from which it follows that
\begin{align}
  \bar{E}_j \simeq \bar{E} + \bar{\xi} \frac{m_j^2}{2\bar{E}} \,, \qquad\quad
  \bar{p}_j \simeq \bar{E} - (1 - \bar{\xi}) \frac{m_j^2}{\bar{E}} \,, \qquad\quad
  \bar{v}_j \simeq 1 - \frac{m_j^2}{2 \bar{E}^2} \,,          \label{eq:qm-rel-Epv}
\end{align}
where
\begin{align}
  \bar{E} \equiv \frac{1}{2} (E_{S,0} + E_{D,0})  \qquad\text{and}\qquad
  1 - \bar{\xi} \equiv \frac{\bar{E}}{2}
      \bigg( \frac{1-\xi_S}{E_{S,0}} + \frac{1-\xi_D}{E_{D,0}} \bigg) \,.
\end{align}
In these expressions, $E_{S,0}$ and $E_{D,0}$ are the mean energies for the
case of massless neutrinos, and $\xi_S$, $\xi_D$ are constant parameters
determined by the properties of the source and the detector, respectively.
These parameters can be calculated only in an explicit treatment of the
neutrino production and detection processes. For conventional neutrino sources,
$\xi_S$ and $\xi_D$ are of $\mathcal{O}(1)$, but for \MB\ neutrinos, the
energies associated with different neutrino mass eigenstates have to coincide
within the line widths $\gamma_S$ and $\gamma_D$, so that $\xi_S$, $\xi_D$ and
$\bar{\xi}$ must be extremely small in this case. Indeed, we will see below that,
in order to reproduce our QFT result \eqref{eq:qft-Gamma}, we have to take
$\xi_S = \xi_D = 0$.

Plugging \eqref{eq:qm-rel-Epv} into \eqref{eq:Jjk-1}, neglecting terms
containing the small product $\Delta m_{jk}^2 (E_{S,0} - E_{D,0}) / \bar{E}^2$
and, in the denominator, also neglecting terms of order $\tilde{\gamma} m_j^2 /
\bar{E}^2$, we obtain
\begin{align}
  J_{jk} &= \frac{A^{(S)}_{jk} - A^{(D)}_{jk}}
                 {  \frac{1}{2}(\gamma_D - \gamma_S)
                  + i (E_{S,0} - E_{D,0})
                  - i \bar{\xi} \Delta m_{jk}^2 / 2 \bar{E}}
  \label{eq:Jjk-2}
\end{align}
with the oscillation and coherence terms abbreviated as
\begin{align}
  A^{(B)}_{jk}
    &= \exp\bigg[
             - i \frac{\Delta m_{jk}^2 L}{2 \bar{E}}
             - \frac{|\Delta m_{jk}^2| \gamma_B L}{4 \bar{E}^2}
           \bigg]
                                                      \nonumber\\
    &\equiv
       \exp\bigg[
             - 2\pi i \frac{L}{L^{\rm osc}_{jk}}
             - \frac{L}{L^{\rm coh}_{B,jk}}
           \bigg] \,.
  \label{eq:qm-Lorentz-Lcoh}
\end{align}
for $B = \{ S, D \}$.  Note that the $A^{(B)}_{jk}$ are identical to the
quantities of the same name defined in eq.~\eqref{eq:qft-abbrev-A}, up to the
replacement of $E_{B,0}$ by $\bar{E}$, which leads to corrections of
$\mathcal{O}( \Delta m_{jk}^2 (E_{S,0} - E_{D,0}) / \bar{E}^2)$.  Since we have
neglected terms of this order in the derivation of \eqref{eq:Jjk-2}, we should
for consistency also neglect them here. The full expression for $\Gamma$ is
\begin{align}
  \Gamma
  &= \frac{\Gamma_0 B_0}{4\pi L^2} \, Y_S Y_D
     \frac{1}{2\pi} \sum_{j,k} |U_{ej}|^2 |U_{ek}|^2 \,
     \exp\bigg[\! -\frac{2 \bar{E}^2 - m_j^2 - m_k^2}{2 \sigma_p^2} \bigg]
     \gamma_S \gamma_D
                                                      \nonumber\\
  &\quad\cdot
     \bigg[
         E_{S,0} - E_{D,0}
       - m_j^2 \bigg( \frac{1-\xi_S}{2 E_{S,0}} - \frac{1-\xi_D}{2 E_{D,0}} \bigg)
       - \frac{i(\gamma_S + \gamma_D)}{2}
     \bigg]^{-1}
                                                      \nonumber\\
  &\quad\cdot
     \bigg[
         E_{S,0} - E_{D,0}
       - m_k^2 \bigg( \frac{1-\xi_S}{2 E_{S,0}} - \frac{1-\xi_D}{2 E_{D,0}} \bigg)
       + \frac{i(\gamma_S + \gamma_D)}{2}
     \bigg]^{-1}
                                                      \nonumber\\
  &\quad\cdot
    \Bigg\{
        \frac{A^{(S)}_{jk}}{\gamma_S + i \xi_S \frac{\Delta m_{jk}^2}{2 E_{S,0}}}
      + \frac{A^{(D)}_{jk}}{\gamma_D - i \xi_D \frac{\Delta m_{jk}^2}{2 E_{D,0}}}
      + \frac{A^{(S)}_{jk} - A^{(D)}_{jk}}
             {    \frac{1}{2}(\gamma_D - \gamma_S)
              \pm i (E_{S,0} - E_{D,0})
              -   i \bar{\xi} \frac{\Delta m_{jk}^2}{2 \bar{E}}}
    \Bigg\} \,.
  \label{eq:qm-Gamma-2}
\end{align}
In the last term, the upper sign applies to the case $\Delta m_{jk}^2 > 0$,
while the lower one is for $\Delta m_{jk}^2 < 0$. As discussed above, $\xi_S$
and $\xi_D$ are very small for \MB\ neutrinos. If we neglect them completely,
$\Gamma$ simplifies to
\begin{align}
  \Gamma
  &= \frac{\Gamma_0 B_0}{4\pi L^2} \, Y_S Y_D
     \frac{1}{2\pi} \sum_{j,k} |U_{ej}|^2 |U_{ek}|^2 \,
     \exp\bigg[\! -\frac{2 \bar{E}^2 - m_j^2 - m_k^2}{2 \sigma_p^2} \bigg]
     \frac{1}
          {(E_{S,0} - E_{D,0})^2 + \frac{1}{4} (\gamma_S + \gamma_D)^2}
                                                      \nonumber\\
  &\cdot
    \Bigg[
        \frac{\gamma_S + \gamma_D}{2} (A^{(S)}_{jk} + A^{(D)}_{jk})
      - \frac{1}{2}
        \frac{(A^{(S)}_{jk} - A^{(D)}_{jk})
              \big[
                (E_{S,0} - E_{D,0})(\gamma_S - \gamma_D) \pm i \frac{(\gamma_S + \gamma_D)^2}{2}
              \big]}
             {E_{S,0} - E_{D,0} \pm i \frac{\gamma_S - \gamma_D}{2}}
    \Bigg] \,.
  \label{eq:qm-Gamma-3}
\end{align}
This equation is identical to our QFT result, eq.~\eqref{eq:qft-Gamma} (within
the approximations made in the two approaches). In particular, we find the same
oscillation, coherence, and resonance terms.

\section{Discussion and conclusions}
\label{sec:discussion}

Let us now summarize and discuss our results. In the first part of this paper,
we have used quantum field theoretical techniques to derive the rate $\Gamma$
of \MB\ neutrino emission, propagation, and absorption
(eq.~\eqref{eq:qft-Gamma}). For the first time, we have explicitly included the
effect of homogeneous line broadening due to fluctuating electromagnetic fields
in the solid state crystals forming the source and the detector. We have confirmed
the expectation from ref.~\cite{Akhmedov:2008jn}
that the resulting formula for $\Gamma$ agrees precisely with the one obtained
in \cite{Akhmedov:2008jn} for the case of inhomogeneous line broadening
caused by crystal defects and impurities.  In particular, we have confirmed
that, also for homogeneous line broadening, $\Gamma$ has a Breit-Wigner-like
resonance structure, and contains oscillation, localization, and coherence
exponentials. Moreover, our formula accounts for the suppression of recoilless
emission and absorption processes compared to their non-recoilless counterparts
through a generalized Lamb-\MB\ factor.  We have also noted that in realistic
experiments the localization and decoherence terms are irrelevant and may be
set equal to unity.  The localization term enforces the condition that the
quantum mechanical delocalization of the neutrino source and detector have to
be small compared to the oscillation lengths for oscillations to take place, a
condition that is easily fulfilled in any oscillation experiment. The
decoherence term, on the other hand, accounts for the possibility of wave
packet separation due to the different group velocities associated with
different neutrino mass eigenstates, but also this does not happen in
terrestrial experiments.

We have then proceeded to a derivation of $\Gamma$ in a quantum mechanical
approach, in which the neutrino is described by a Lorentzian wave packet of the
form \eqref{eq:qm-Lorentz-1}. We have arrived at eq.~\eqref{eq:qm-Gamma-3},
which coincides with the QFT result \eqref{eq:qft-Gamma}. However, since the
neutrino production and detection processes, which involve particle creation
and annihilation, cannot be described in QM, the \MB\ neutrino production rate
as well as the detection cross section had to be put in by hand. Also, the
properties of the neutrino wave packets (shape, width, central momenta) had to
be chosen in an ad hoc way instead of emerging naturally from the formalism or
being related to properties of the source and the detector.  Once the
appropriate choices for these parameters are made, the Breit-Wigner-shaped
resonance factor as well as the oscillation and decoherence terms can be
derived.  The correct Lamb-\MB\ and localization factors are obtained only if
suitably chosen phenomenological weighting factors $f_{jS}$, $f_{jD}$ for the
different neutrino mass eigenstates are introduced in the neutrino wave
function to account for the tiny dependence of the emission and absorption
probabilities on the neutrino mass.

As expected, $\Gamma$ factorizes into the emitted neutrino flux, a transition
probability $\mathcal{P}(L)$, and the detection cross section.  While in the QM
approach, this property is introduced as an assumption in eq.\ \eqref{eq:fact},
it emerges naturally in QFT. The reason is that for large propagation distance
$L$ off-shell effects become negligible, and according to the Grimus-Stockinger
theorem eq.~\eqref{eq:Grimus} the propagator then reduces to the exponential
phase factor $\exp(i p L)$ (with $p$ being the modulus of the neutrino
momentum), which is also used in QM to describe the spatial evolution of
particles.

In conclusion, we have shown that the QM approach to \MB\ neutrino
oscillations, in which the production, propagation, and detection processes are
treated separately, is able to reproduce the results obtained in the QFT
approach, in which these processes are a priori considered as a single entity
and their factorization emerges as a result. In general, the framework of QFT
is significantly more robust because it does not require any assumptions on the
neutrino wave function, whose parameters are instead automatically determined
from the much less ambiguous properties of the neutrino source and the
detector. For example, homogeneous and inhomogeneous line broadening are easy
to implement in QFT (see sec.~\ref{sec:qft} and ref.~\cite{Akhmedov:2008jn}),
while in QM, they have to be accounted for by choosing appropriate wave packet
widths. Also, the emission rate, the detection cross section, and the Lamb-\MB\
factor cannot be predicted in QM and have to be put in by hand. On the other
hand, the QM approach can give a better physical understanding of the origin of
oscillation, decoherence, and resonance phenomena once all free parameters are
chosen appropriately, e.g.\ by matching with the QFT result.

\section*{Acknowledgments}

It is a pleasure to thank Evgeny Akhmedov, Samoil Bilenky, Franz von
Feilitzsch, Manfred Lindner, and Walter Potzel for interesting and helpful
discussions.  This work was in part supported by the Transregio
Sonderforschungsbereich TR27 ``Neutrinos and Beyond'' der Deutschen
Forschungsgemeinschaft. The author would also like to acknowledge support from
the Studienstiftung des Deutschen Volkes.

\begin{center}
  \rule{10cm}{0.25pt}
\end{center}

\end{document}